# Stability boundaries for wrinkling in highly stretched elastic sheets

Qingdu Li[a], Timothy J. Healey[b,*]

[a]Key Laboratory of Industrial Internet of Things & Networked Control, Ministry of Education, Chongqing University of Posts and Telecommunications, Chongqing, 400065, China.
[b]Department of Mathematics, Cornell University, Ithaca, NY, USA
[*]Corresponding author: Tel. +1 607 255 8282. *Email address*: tjh10@cornell.edu

ABSTRACT

We determine stability boundaries for the wrinkling of highly unidirectionally stretched, finely thin, rectangular elastic sheets. For a given fine thickness and length, a *stability boundary* here is a curve in the parameter plane, aspect ratio vs. the macroscopic strain; the values on one side of the boundary are associated with a flat, unwrinkled state, while wrinkled configurations correspond to all values on the other. In our recent work we demonstrated the importance of finite elasticity in the membrane part of such a model in order to capture the correct phenomena. Here we present and compare results for four distinct models: (i) the popular Föppl-von Kármán plate model (FvK), (ii) a correction of the latter, used in our earlier work, in which the approximate 2D Föppl strain tensor is replaced by the exact Green strain tensor, (iii) and (iv): effective 2D finite-elasticity membrane models based on 3D incompressible neo-Hookean and Mooney-Rivlin materials, respectively. In particular, (iii) and (iv) are superior models for elastomers. The 2D nonlinear, hyperelastic models (ii)-(iv) all incorporate the same quadratic bending energy used in FvK. Our results illuminate serious shortcomings of the latter in this problem, while also pointing to inaccuracies of model (ii) – in spite of yielding the correct qualitative phenomena in our earlier work. In each of these, the shortcoming is a due to a deficiency of the membrane part of the model.

________________________________________________________________________________

________________________________________________________________________________

## 1. Introduction

We return in this work to our study of the wrinkling of highly stretched, finely thin, rectangular elastic sheets, cf. Healey, et. al. (2013). The problem concerns the development of transverse wrinkles as the two opposing shorter, clamped ends are pulled apart. The phenomenon is well known and has been studied from many different points of view, e.g., Friedl, et. al. (2000), Cerda et. al. (2002), Jacques and Potier-Ferry (2005), Puntel et. al. (2011). We refer to the introduction to our earlier paper for an overview of some features of those works. In particular, none of these employs a geometrically exact membrane model. We postpone a discussion of the works Taylor, et. al. (2014) and Nayyar, et. al. (2011), (2014), which employ models closely related to ours, until the end of this section.

Two novel ingredients of our paper from 2013 are: (i) a geometrically exact, elastic membrane model perturbed by very small bending stiffness; (ii) the systematic use of global bifurcation/continuation methods via multiple parameters - including the macroscopic strain, membrane thickness, and aspect ratio - to uncover stable wrinkling behavior. For a given fine thickness, two key findings are: (1) Wrinkling occurs only for aspect ratios contained in a bounded interval, i.e., the stretched membrane remains planar without wrinkles if the aspect ratio is either too large or too small. (2) When wrinkling occurs, as the macroscopic strain is continuously increased, wrinkles first initiate (bifurcate) at a non-zero value, reach a very small, maximum amplitude, diminish, and then disappear altogether. Mathematically this corresponds to an *isola-* *center* bifurcation diagram (the nontrivial solution curve "starts" and "stops" at two distinct points on the trivial line), cf. Golubitsky and Schaeffer (1985). In contrast, the popular Föppl-von Kármán (FvK) model, also considered in Healey, et. al. (2013), yields *pitchfork* bifurcation diagrams (like the post-buckling curve of an Euler column) for seemingly all aspect ratios – large and small. Observe that a "pitchfork" here unrealistically implies an ever-increasing wrinkling amplitude as the macroscopic stain is increased.

The model employed in Healey, et. al (2013) is a correction to the FvK model – Green's strain tensor is employed in lieu of the Föppl approximation. In particular, this results in a geometrically exact membrane model characterized by the Saint-Venant Kirchhoff (S-VK) constitutive law, the mathematical tenability of which, in a highly unidirectional tensile state, is analyzed in our paper. Although the model captures the correct qualitative phenomena, the overall deficiency of that constitutive law as a realistic model for elastomers is clear. One goal of this work is to extend our previous approach to superior constitutive models for membrane elasticity. In particular, we employ 2D models based on the 3D incompressible neo-Hookean (NH) and Mooney-Rivlin (MR) materials, cf. Mooney (1940), Müller and Strehlow (2004). Our modelling philosophy here is the same as before, viz, the elastic sheet is idealized as a 2D nonlinearly elastic membrane perturbed by extremely small, linear bending stiffness.

Another motivation for our study comes from the tacit suggestion in Healey, et. al. (2013) (echoed above) that the FvK model predicts eventual wrinkling – for sufficiently large macroscopic strain – for *all* aspect ratios. In fact we demonstrate here that this is not entirely true, although something along those lines indeed holds, as discussed below. In any case, to efficiently address such questions, our investigation here is focused on the computation of *stability boundaries*: For a given fine thickness and length, the latter is defined as a curve of bifurcation points (engendering wrinkling) in the parameter plane spanned by the aspect ratio and the macroscopic strain. The boundary separates parameter values associated with flat, unwrinkled states from those corresponding to wrinkled configurations. As a basis for comparison, we compute the stability boundaries (for two distinct fine thicknesses) employing the four models mentioned above: FvK, S-VK, NH and MR.

The outline of the work is as follows. In Section 2 we formulate the problem. In addition to the FvK model and the S-VK model, both studied in Healey, et. al. (2013), we present the two more accurate models, NH and MR. Following the derivation in Müller and Strehlow (2004), for example, the thin membrane is treated as an incompressible, 3D elastomer. The constraint of incompressibility eliminates the through-thickness stretch, while the constitutively indeterminate pressure is employed to achieve zero tractions on the lateral surfaces of the membrane. All of this results in an effective 2D model. Here we tune the parameter(s) so that shear moduli of the two models agree with that inherent in FvK and S-VK. The same small quadratic bending energy is consistently employed in each of the four models – tuned to the incompressible limit corresponding to Poisson's ratio equal to 0.5. Our justification for this simple choice is the following. All previous studies, both numerical, e.g., Nayyar, et. al. (2011), Healey, et. al. (2013), Taylor, et. al (2014), and experimental, e.g., Zheng (2009), Fehér and Sipos (2014), Nayyar, et. al. (2014), indicate that the maximum wrinkling displacement is of the same order of magnitude as the fine thickness, while the wavelength of the typical wrinkle is two orders of magnitude greater. We comment further on this in the final section of the paper.

In Section 3 we discuss our strategy for the efficient computation of stability boundaries. Due to the reflection symmetry of the problem across the flat configuration, we demonstrate that the tangent operator evaluated at a planar configuration is block-diagonal, comprising two blocks – one corresponding to the out-of-plane scalar displacement field, the other to the in-plane vector displacement field. In this way we can compute a stability boundary via continuation for an inflated system involving the *reduced* equilibrium equations governing planar configurations combined with the zero-eigenvalue problem for the above-mentioned block associated with out-of-plane displacements. From a general point of view, such a system is known to be convergent by Newton's method, cf. Werner and Spence (1984). Instead of the latter, however, we solve the two components of the inflated system separately and consecutively. In particular, we determine the zero-eigenvalue component via a simple mid-point-rule iteration, each iterate requiring the updated solution of the reduced equilibrium equations. This avoids the evaluation of second-derivative operators while enabling efficient computation of stability boundaries. The algorithm is summarized in Section 2.

In Section 4 we present our results for two distinct fine thicknesses. For each thickness we first present stability boundaries for the four models – FvK, S-VK, NH, MR – in terms of aspect ratio vs. macroscopic strain. The curves obtained using the FvK model open to the right, predicting a convex and apparently unbounded region of wrinkling. In contrast, each of the three finite-elasticity membrane models, S-VK, NH and MR, yield closed-curve stability boundaries, with the wrinkling regions inside the curves. For the same thickness, the NH and MR boundaries compare very well; the S-VK curves, although closed and thus qualitatively correct, do not compare well with the two more accurate models. We also provide numerous bifurcation diagrams for the various models, featuring the FvK pitchforks and the isloa-centers coming from the other three models. Finally we illustrate some wrinkled configurations along specific solution diagrams for the two new models, NH, MR, considered in this work. As in Healey, et. al. (2013), we avoid the numerical pitfalls of a near-zero bending stiffness by the simple step of dividing the equations through by the highest power of the thickness and then employing the reciprocal of the squared thickness as an increasing continuation parameter. In particular, this leaves the principal part of the bending operator fully intact, cf. Section2. In this way, we find all bifurcation diagrams accurately and reliably. Using the same methodology from our earlier paper, all nontrivial, wrinkled solutions presented in this work are locally stable in the sense that the total potential energy is verified to be a local minimum there. As verified in Healey et. al. (2013), there is also a plethora of unstable wrinkled solutions. In Section 5 we make some final remarks.

We now discuss the works employing models closely related to ours considered here. In Taylor, et. al. (2014) the finite-strain Koiter model is employed, cf. Ciarlet (2005), Steigmann (2013). As pointed out in Section 2 of this work, the only difference between that and the S-VK model is that the former employs the exact relative curvature tensor instead of its linearization in the bending energy. In other words, the 2D geometrically exact membrane part is common to both models; its apparent inaccuracy in predicting stability boundaries is indicated in Section 4. In Taylor, et. al. (2014), the problem is formulated dynamically including inertial effects, in a highly overdamped environment. Accordingly, for fixed parameter values, the numerical solutions of initial-value problems, with "randomly" chosen initial conditions, rapidly decay to apparently stable equilibria. In this way several wrinkled states, for a specific set of parameters, are obtained and reported in that work. No systematic study of the problem, e.g., bifurcation diagrams, dependence on other parameters, etc., is considered. Other types of interesting wrinkling problems are addressed in the paper as well.

The paper Nayyar, et. al. (2015) is a follow-up to their earlier work Nayyar, et. al. (2011). The latter presents results using the commercial code ABACUS. In particular, a finite-deformation, hyperelastic shell element, parametrized by an extremely small thickness, is employed. The membrane part of that model is identical to that of the NH model used here. Using an ad-hoc imperfection-perturbation approach, approximate bifurcation diagrams are obtained, which agree with the trend (2) discussed in the second paragraph of this section. Results for various fixed thicknesses and aspect ratios are presented. No concerns about numerical reliability are expressed. No stability information about solutions is reported. The more recent paper presents interesting experimental results, again verifying the trend (2), along with measured wrinkled profiles. In particular, these results indicate a rate dependence on the diminishing amplitude of wrinkles in the large-strain state. The paper concludes with a numerical study of visco-elastic membranes, again employing ABACUS, in order to understand this phenomenon.

**2. Formulation**

Let $\{\mathbf{e}_1, \mathbf{e}_2, \mathbf{e}_3\}$ denote a fixed orthonormal basis for the Euclidean vector space $\mathbb{E}^3$. We assume that the stress-free, undeformed rectangular sheet, denoted $\Omega$, lies in the plane $\mathbb{E}^2 = span\{\mathbf{e}_1, \mathbf{e}_2\}$; $\Omega = \{\mathbf{x} := x_\alpha \mathbf{e}_\alpha : x_1 \in (0, L), x_2 \in (0, W)\}$, cf. Figure 1. Here and throughout we employ summation convention, with repeated Latin indices summing from 1 to 3, and repeated Greek indices summing from 1 to 2. We denote the *deformation* of the sheet, $\mathbf{f} : \Omega \to \mathbb{E}^3$, via

$$\mathbf{f}(\mathbf{x}) := \mathbf{x} + \mathbf{u}(\mathbf{x}) + w(\mathbf{x})\mathbf{e}_3, \tag{1}$$

where $\mathbf{u} : \Omega \to \mathbb{E}^2$ and $w : \Omega \to \mathbb{R}$ denote the in-plane displacement and the out-of-plane displacement component, respectively. As illustrated in Figure 1, we fix the sheet on the left side of the boundary at $x_1 = 0$, and prescribe the displacement on the right, while the top and bottom remain free. The prescribed geometric boundary conditions are thus

$$\begin{aligned}\mathbf{u}(0, x_2) = \mathbf{0},\ \mathbf{u}(L, x_2) \equiv \varepsilon L \mathbf{e}_1, \\ w(0, x_2) = w(L, x_2) = 0,\ 0 \leq x_2 \leq W,\end{aligned} \tag{2}$$

where $\varepsilon \geq 0$ is the *macroscopic strain*.

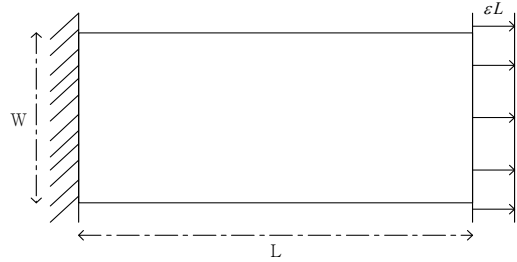

Figure 1. Schematic diagram of a stretched elastic sheet

The *deformation gradient* takes the form

$$\mathbf{F} := \nabla \mathbf{f} = \mathbf{I} + \nabla \mathbf{u} + \mathbf{e}_3 \otimes \nabla w, \tag{3}$$

where $\nabla(\cdot)$ denotes the surface gradient, viz., $\nabla \mathbf{f} := f_{i,\beta} \mathbf{e}_i \otimes \mathbf{e}_\beta$, $\nabla \mathbf{u} := u_{\alpha,\beta} \mathbf{e}_\alpha \otimes \mathbf{e}_\beta$, $\mathbf{I} := \mathbf{e}_\alpha \otimes \mathbf{e}_\alpha$ is the identity tensor on $\mathbb{E}^2$, $f_{i,\beta}$ denotes the partial derivative of $f_i$ with respect to $x_\beta$, etc., and "$\otimes$" is the tensor product. The *membrane* right Cauchy-Green *strain* is the symmetric tensor given by

$$\mathbf{C} = \mathbf{F}^T \mathbf{F} = \mathbf{I} + \nabla \mathbf{u} + \nabla \mathbf{u}^T + \nabla \mathbf{u}^T \nabla \mathbf{u} + \nabla w \otimes \nabla w, \tag{4}$$

which we presume to be positive definite. The *unit normal* to the deformed surface $\Sigma := \mathbf{f}(\Omega)$ at $\mathbf{y} = \mathbf{f}(\mathbf{x})$ is then

$$\mathbf{n} = \frac{\mathbf{F}\mathbf{e}_1 \times \mathbf{F}\mathbf{e}_2}{\sqrt{\det \mathbf{C}}}. \tag{5}$$

The *second gradient* of the deformation is

$$\nabla^2 \mathbf{f} := \nabla \circ \nabla \mathbf{f} = \nabla \mathbf{F} = \nabla^2 \mathbf{u} + \mathbf{e}_3 \otimes \nabla^2 w, \tag{6}$$

Observe that (at fixed $\mathbf{x} \in \Omega$) both $\nabla^2 \mathbf{f}(\mathbf{x})$ and $\nabla^2 \mathbf{u}(\mathbf{x})$ are symmetric-bilinear but with different ranges: $\nabla^2 \mathbf{f}(\mathbf{x})[\mathbf{a}, \mathbf{b}] = \nabla^2 \mathbf{f}(\mathbf{x})[\mathbf{b}, \mathbf{a}] \in \mathbb{E}^3; \nabla^2 \mathbf{u}(\mathbf{x})[\mathbf{a}, \mathbf{b}] = \nabla^2 \mathbf{u}(\mathbf{x})[\mathbf{b}, \mathbf{a}] \in \mathbb{E}^2$ for all $\mathbf{a}, \mathbf{b} \in \mathbb{E}^2$. $\nabla^2 w(\mathbf{x})$ is the *Hessian* of $w$ at $\mathbf{x}$. The *relative curvature* of $\Sigma$ is defined by

$$\mathbf{K} := -\mathbf{n} \cdot \nabla^2 \mathbf{f}. \tag{7}$$

Indeed, if we define $\mathbf{a}_\alpha := \mathbf{f}_{,\alpha}$ then $\mathbf{F} = \mathbf{a}_\alpha \otimes \mathbf{e}_\alpha$ and $\nabla^2 \mathbf{f} = \mathbf{a}_{\alpha,\beta} \otimes \mathbf{e}_\alpha \otimes \mathbf{e}_\beta$. Definition (7) now reads $\mathbf{K} = -(\mathbf{n} \cdot \mathbf{a}_{\alpha,\beta}) \mathbf{e}_\alpha \otimes \mathbf{e}_\beta$, where $\mathbf{n}_{,\alpha} \cdot \mathbf{a}_\beta = -\mathbf{n} \cdot \mathbf{a}_{\beta,\alpha} = -\mathbf{n} \cdot \mathbf{a}_{\alpha,\beta}$ are the components of the second fundamental form on $\Sigma$.

We consider a class of objective stored-energy functions of the form

$$W(\mathbf{C},\mathbf{K}) = \Psi(\mathbf{K}) + \Phi(\mathbf{C}), \tag{8}$$

where $\Psi$ and $\Phi$ account for bending and membrane effects, respectively. For the latter we treat the membrane as a thin 3D, incompressible neo-Hookean or Mooney-Rivlin solid. Following the derivation in Müller and Strehlow (2004), the 3D incompressibility condition is used in tandem with the pressure, resulting in:

$$\text{Neo-Hookean: } \Phi(\mathbf{C}) = \frac{Eh}{6}\left[ tr\mathbf{C} + (\det \mathbf{C})^{-1} - 3 \right], \tag{9}$$

Mooney Rivlin:

$$\Phi(\mathbf{C}) = \frac{Eh}{6.6}\left\{ \left[ tr\mathbf{C} + (\det \mathbf{C})^{-1} - 3 \right] + 0.1\left[ tr\mathbf{C}(\det \mathbf{C})^{-1} + \det \mathbf{C} - 3 \right] \right\}, \tag{10}$$

where "$E$" denotes Young's modulus and "$h$" the membrane thickness. For purposes of the forthcoming comparison with our previous results from Healey, et. al. (2013), we also summarize two other constitutive models for the 2D membrane. Recall that *Green's strain* tensor is given by

$$\mathbf{E} := (\mathbf{C} - \mathbf{I})/2 = (\nabla \mathbf{u} + \nabla \mathbf{u}^T + \nabla \mathbf{u}^T \nabla \mathbf{u} + \nabla w \otimes \nabla w)/2, \tag{11}$$

while the *Föppl strain* tensor ignores the quadratic term $\nabla \mathbf{u}$ in (11), viz.,

$$\mathbf{H} := (\mathbf{C} - \nabla \mathbf{u}^T \nabla \mathbf{u} - \mathbf{I})/2 = (\nabla \mathbf{u} + \nabla \mathbf{u}^T + \nabla w \otimes \nabla w)/2. \tag{12}$$

The 2D membrane strain energy is then defined (and designated) using these two strain measures via

$$\Phi(\mathbf{C}) := \frac{Eh}{3}\left[ (tr\mathbf{E})^2 + \mathbf{E} \cdot \mathbf{E} \right], \text{ Saint-Venant Kirchhoff;}$$

$$\Phi(\mathbf{C}) := \frac{Eh}{3}\left[ (tr\mathbf{H})^2 + \mathbf{H} \cdot \mathbf{H} \right], \text{ Föppl.} \tag{13}$$

where we have used $\nu = 0.5$ for Poisson's ratio. Each of the stored energy functions (9), (10) and (13)$_{1,2}$ are tuned to the same shear modulus, viz., $G = E/3$.

For the bending energy density, we assume isotropy and invariance under reflection, $w \to -w$. Retaining only quadratic terms, we adopt the Kirchhoff energy (for $\nu = 0.5$):

$$\Psi(\mathbf{K}) = \frac{Eh^3}{36}\left[ (tr\mathbf{K})^2 + \mathbf{K} \cdot \mathbf{K} \right]. \tag{14}$$

Finally, for small transverse displacements and gradients, $w, |\nabla w|$, we make the approximation $\mathbf{n} \cong \mathbf{e}_3$. In view of (6) and (7), we then have

$$\mathbf{K} \cong -\nabla^2 w, \tag{15}$$

which we henceforth employ in (14). Observe that (13)$_2$, cf. (12), together with (14), (15) comprise the classical Föppl-von Kármán model, denoted FvK; (13)$_1$, cf. (11), with (14), (15) is the model employed in Healey, et. al. (2013), henceforth denoted S-VK. The models (9) and (10), each combined with (14), (15) are denoted NH and MR, respectively. We remark further that (13)$_1$ combined with (14) - without the approximation (15) - is usually called the Koiter model, cf. Ciarlet (2005), Stiegmann (2013).

The absence of body forces and surface tractions allow us to normalize the stored energy as

$$\bar{\Phi}(\mathbf{C}) := \frac{9}{Eh^3}\Phi(\mathbf{C}),$$
$$\bar{\Psi}(\mathbf{K}) := \frac{9}{Eh^3}\Psi(\mathbf{K}),$$
(16)

without altering the problem at hand, where the total (scaled) potential energy is now given by

$$U = \int_{\Omega}[\bar{\Phi}(\mathbf{C}) + \bar{\Psi}(\mathbf{K})]dx. \tag{17}$$

Next we define the *stress* and *stress couple*

$$\mathbf{N} := 2\frac{\partial\bar{\Phi}}{\partial\mathbf{C}}(\mathbf{C}) \text{ and } \mathbf{M} := \frac{\partial\bar{\Psi}}{\partial\mathbf{K}}(\mathbf{K}), \tag{18}$$

respectively. Stationary potential energy then yields the weak form of the equilibrium equations:

$$\delta U = \int_{\Omega}\left[\kappa[(\mathbf{I}+\nabla\mathbf{u})\mathbf{N}\cdot\nabla\boldsymbol{\eta} + (\mathbf{N}\nabla w)\cdot\nabla\zeta] - \mathbf{M}\cdot\nabla^2\zeta\right]d\mathbf{x} = 0,$$
$$\kappa := 1/h^2,$$
(19)

for all *admissible variations*, i.e., for all smooth fields $\boldsymbol{\eta}:\Omega\to\mathbb{E}^2$ and $\zeta:\Omega\to\mathbb{R}$ vanishing along the ends $x_1 = 0$ and $x_1 = L$, cf. (2). A formal integration by parts yields the partial differential equations of equilibrium, listed here for clarity:

$$\nabla\cdot[(\mathbf{I}+\nabla\mathbf{u})\mathbf{N}] = \mathbf{0},$$
$$\Delta^2 w - \kappa\nabla\cdot(\mathbf{N}\nabla w) = 0 \text{ in } \Omega.$$
(20)

Assuming fixed $L$, we see that our problem is characterized by two non-negative parameters, viz., the macroscopic strain $\varepsilon$, cf. (2), and the *thickness parameter* $\kappa = 1/h^2$, cf. (19). Another important parameter for the purposes of our study is the aspect ratio, which we introduce as follows. Defining $\beta := L/W$, we then rescale the independent variable $\mathbf{e}_2\cdot\mathbf{x} = x_2 \to x_2/\beta$, and henceforth work on the fixed square domain $\tilde{\Omega} = (0,L)\times(0,L)$. Define the diagonal transformation

$$\mathcal{I}_{\beta} := \mathbf{e}_1\otimes\mathbf{e}_1 + \beta\mathbf{e}_2\otimes\mathbf{e}_2. \tag{21}$$

Then the gradients appearing in our formulation above are replaced as follows:

$$\nabla w \to \mathcal{I}_{\beta}\nabla w,$$
$$\nabla\mathbf{u} \to \nabla\mathbf{u}\mathcal{I}_{\beta},$$
$$\nabla^2 w \to \mathcal{I}_{\beta}\nabla^2 w\mathcal{I}_{\beta}.$$
(22)

In what follows we employ a finite-element method to numerically solve (20). Accordingly we return to the (rescaled) weak form (19):

$$\mathcal{F}_1(\mathbf{u},w;\varepsilon,\beta,\kappa) := \int_{\tilde{\Omega}}\kappa(\mathbf{I}+\nabla\mathbf{u}\mathcal{I}_{\beta})\mathbf{N}\cdot\nabla\boldsymbol{\eta}d\mathbf{x} = \mathbf{0} \text{ for all } \boldsymbol{\eta},$$
$$\mathcal{F}_2(\mathbf{u},w;\varepsilon,\beta,\kappa) := \int_{\tilde{\Omega}}\left[\kappa(\mathbf{N}\mathcal{I}_{\beta}\nabla w)\cdot\nabla\zeta - \mathcal{I}_{\beta}\mathbf{M}\mathcal{I}_{\beta}\cdot\nabla^2\zeta\right]d\mathbf{x} = 0 \text{ for all } \zeta,$$
(23)

where we have incorporated (22). In view of (20), note that (23)$_1$ is vector-valued, representing the in-plane equilibrium conditions, while (23)$_2$, expressing moment-balance, is scalar valued. The abstract representations of the weak form of the equilibrium equations, viz.

$$\mathcal{F}(\mathbf{u}, w; \varepsilon, \beta, \kappa) := (\mathcal{F}_1(\mathbf{u}, w; \varepsilon, \beta, \kappa), \mathcal{F}_2(\mathbf{u}, w; \varepsilon, \beta, \kappa)) = (\mathbf{0}, 0), \tag{24}$$

is especially convenient for our development in Section 3.

## 3. Computation of Stability Boundaries

In this section we present an efficient strategy for computing wrinkling-stability boundaries, the latter of which we define as follows: For a given fine thickness $h = 1/\sqrt{\kappa}$ and length $L$, a *wrinkling-stability boundary* is a curve in the $\varepsilon$-$\beta$ plane separating parameter regions associated with unwrinkled, planar configurations from those representing wrinkled, non-planar configurations.

We begin with the symmetries inherent in (24), which play a key role in our computational strategy. First consider the reflection $w \to -w$, under which (4) and (15) imply $\mathbf{C} \to \mathbf{C}$ and $\mathbf{K} \to -\mathbf{K}$, respectively. By virtue of (14), $\Psi$ is an even function of $\mathbf{K}$, and from (18), it follows that $\mathbf{N} \to \mathbf{N}$ and $\mathbf{M} \to -\mathbf{M}$ under reflection. Finally, (23) combined with these considerations yield

$$\begin{aligned}\mathcal{F}_1(\mathbf{u}, -w; \varepsilon, \beta, \kappa) &= \mathcal{F}_1(\mathbf{u}, w; \varepsilon, \beta, \kappa), \\ \mathcal{F}_2(\mathbf{u}, -w; \varepsilon, \beta, \kappa) &= -\mathcal{F}_2(\mathbf{u}, -w; \varepsilon, \beta, \kappa).\end{aligned} \tag{25}$$

We note that (25)$_2$ leads immediately to

$$\mathcal{F}_2(\mathbf{u}, 0; \varepsilon, \beta, \kappa) \equiv 0, \tag{26}$$

i.e., the bending equilibrium equation is identically satisfied, cf. (20)$_2$. Hence, planar configurations are governed solely by the reduced problem

$$\mathcal{F}_1(\mathbf{u}, 0; \varepsilon, \beta, \kappa) = \mathbf{0}. \tag{27}$$

The reflection symmetries (25) further imply that the total derivative or *linearization* of (24) at an arbitrary planar configuration ($w \equiv 0$) is block diagonal. To see this, first note that the linearization has the general block form (suppressing the dependence on parameters)

$$D\mathcal{F}(\mathbf{u}, w) = \left[\begin{array}{c|c} D_u \mathcal{F}_1(\mathbf{u}, w) & D_w \mathcal{F}_1(\mathbf{u}, w) \\ \hline D_u \mathcal{F}_2(\mathbf{u}, w) & D_w \mathcal{F}_2(\mathbf{u}, w) \end{array}\right], \tag{28}$$

where $D_u(\cdot)$ and $D_w(\cdot)$ denote total (partial) derivatives. Next we take an arbitrary directional derivative of (25)$_1$ with respect to $w$ in the direction of a smooth, scalar-valued function $\phi$: replace $w$ by $w + \alpha\phi$ in (25)$_1$, for a scalar $\alpha$, compute the derivative with respect to $\alpha$ via the chain rule, and then evaluate at $\alpha = 0$ and $w \equiv 0$, all leading to

$$\begin{aligned}-D_w \mathcal{F}_1(\mathbf{u}, 0)\phi &= D_w \mathcal{F}_1(\mathbf{u}, 0)\phi \text{ for all } \phi, \\ \Rightarrow D_w \mathcal{F}_1(\mathbf{u}, 0) &\equiv 0.\end{aligned} \tag{29}$$

Similarly, an arbitrary directional derivative of (25)$_2$ with respect to $\mathbf{u}$ in the direction of a smooth, vector-valued function $\xi$, yields

$$D_u \mathcal{F}_2(\mathbf{u},0)\xi = -D_u \mathcal{F}_2(\mathbf{u},0)\xi \text{ for all } \xi,$$
$$\Rightarrow D_u \mathcal{F}_2(\mathbf{u},0) \equiv 0. \tag{30}$$

Finally (29), (30) and the evaluation of (28) at $w \equiv 0$ delivers the block-diagonal linearization:

$$D\mathcal{F}(\mathbf{u},0) = \left[\begin{array}{c|c} D_u \mathcal{F}_1(\mathbf{u},0) & 0 \\ \hline 0 & D_w \mathcal{F}_2(\mathbf{u},0) \end{array}\right]. \tag{31}$$

It's worth noting that each of the operators $D\mathcal{F}(\mathbf{u},w), D_u\mathcal{F}_1(\mathbf{u},0)$ and $D_w\mathcal{F}_2(\mathbf{u},0)$ is self-adjoint. This follows from (17) and (19); each is a certain second derivative of the energy $U$. For example, $D_w\mathcal{F}_2(\mathbf{u},0) \equiv D_w^2 U(\mathbf{u},0)$, where $D_w^2(\cdot)$ denotes the second total derivative with respect to "$w$".

Now the planar system ($w \equiv 0$) will suffer an instability precisely at configurations $\mathbf{u}$ for which the linearization is singular, and (31) shows that "in-plane" instabilities, associated with the top-left block, are decoupled from wrinkling instabilities, associated with the bottom-right block. In fact, we do not anticipate the former and concentrate on the latter. In what follows, we assume that a given fine thickness $h$ is fixed, and we suppress the explicit appearance of $\kappa = 1/h^2$ in (24); at the same time we re-introduce the dependence on the other two parameters. In view of (27) and (31), we say that $(\mathbf{u}, \varepsilon, \beta)$ is a point of *wrinkling instability* if

$$\mathcal{F}_1(\mathbf{u},0;\varepsilon,\beta) = 0,$$
$$D_w\mathcal{F}_2(\mathbf{u},0;\varepsilon,\beta)\varphi = 0, \tag{32}$$

for some *null vector* $\varphi \neq 0$. The locus of all points $(\varepsilon, \beta)$ such that $(\mathbf{u}, \varphi; \varepsilon, \beta)$ satisfy (32) comprise the stability boundary.

If one of the parameters is fixed, say $\beta$ =const., and if we add a normalizing condition for the length of the null vector $\varphi$, then the resulting inflated system (32) is regular at a symmetry-breaking bifurcation point, cf. Werner and Spence (1984). However, its implementation via Newton's method requires the evaluation of higher derivatives of $\mathcal{F}_2$, which we wish to avoid. Instead we solve (32)₁ by continuation and adjust (32)₂ via a simple mid-point rule as follows.

Suppose that a solution point of (32) is known, say, $(\mathbf{u}_o, \varphi_o; \varepsilon_o, \beta_o)$. We first increment one of the parameters, say:

$$\beta = \beta_1 := \beta_o + \Delta\beta \text{ with } \varepsilon = \varepsilon_o \text{ fixed}, \tag{33}$$

where $\Delta\beta$ is sufficiently small. Next we update via Newton's method to a solution $\mathbf{u}_1$ of (32)₁, viz.,

$$\mathcal{F}_1(\mathbf{u}_{(1)}, 0; \varepsilon_o, \beta_1) = 0. \tag{34}$$

We then update (32)₂ for the smallest eigenvalue, denoted $\delta_{(1)}$ (positive or negative):

$$D_w\mathcal{F}_2(\mathbf{u}_{(1)}, 0; \varepsilon_o, \beta_1)\varphi_{(1)} = \delta_{(1)}\varphi_{(1)}. \tag{35}$$

Next increment

$$\varepsilon_{(1)} = \varepsilon_o + \Delta\varepsilon_{(1)} \text{ with } \beta = \beta_1, \tag{36}$$

update a solution of (32)₁,

$$\mathcal{F}_1(\mathbf{u}_{(2)}, 0; \varepsilon_{(1)}, \beta_1) = 0, \tag{37}$$

and then update (32)₂ for the smallest eigenvalue:

$$D_w\mathcal{F}_2(\mathbf{u}_{(2)}, 0; \varepsilon_{(1)}, \beta_1)\varphi_{(2)} = \delta_{(2)}\varphi_{(2)}. \tag{38}$$

If $\delta_{(1)}$ and $\delta_{(2)}$ have the same sign, we repeat (36)-(38) with a larger increment $\Delta\varepsilon_{(1)}$, resulting in different signs. Either way, presuming different signs, we set

$$\varepsilon_{(2)} = \varepsilon_{(1)} + \Delta\varepsilon_{(2)}; \quad \Delta\varepsilon_{(2)} = \frac{\delta_{(2)}}{\delta_{(1)} - \delta_{(2)}} \Delta\varepsilon_{(1)}. \tag{39}$$

We repeat the process (35)-(39) "$N$" steps, in the obvious way, until the smallest eigenvalue of $D_w \mathcal{F}_2(\mathbf{u}_{(N+1)}, 0; \varepsilon_{(N)}, \beta_1)$ is sufficiently small. Finally set

$$(\mathbf{u}_1, 0; \varepsilon_1, \beta_1) := (\mathbf{u}_{(N+1)}, 0; \varepsilon_{(N)}, \beta_1), \tag{40}$$

and return to (33). We proceed in this manner via the algorithm (33)-(40) to compute the entire stability boundary connected to the initial solution point $(\mathbf{u}_o, \varphi_o; \varepsilon_o, \beta_o)$. We point out that the roles of the parameters, $\varepsilon$ and $\beta$, can and sometimes must be reversed above in order to efficiently compute the entire boundary.

## 4. Results

As in Healey, et. al. (2013), we analyze one-quarter of the sheet, $(0, L/2) \times (0, L/2)$, enforcing reflection-symmetric conditions along $x_1 = L/2$ and anti-symmetric conditions along $x_2 = L/2$. We also could have considered reflection-symmetric conditions along $x_2 = L/2$. But as noted in our earlier paper, this leads to bifurcation diagrams that are identical to those satisfying anti-symmetry along $x_2 = L/2$; this is due to apparent "asymptotic" symmetry for very small thicknesses. In any case, we fix $L = 20$ for the remainder of this work, and present results for thicknesses $h = 0.01$, and $h = 0.005$ (all, say, in centimeters) in what follows. The numerical implementation of (24) and (32) is identical to that employed in Healey, et. al. (2013), and we refer to that work for the details. In particular, we discretize via a uniform rectangular grid, employing 4-node rectangular conformal finite elements as follows: Each corner of a given rectangular element possesses 6 degrees of freedom corresponding to the values of $u_1, u_2, w, \partial w/\partial x_1, \partial w/\partial x_2$ and $\partial^2 w/\partial x_1 \partial x_2$ there. As established in our earlier work, we use a reliable 70 (length) × 170 (width) uniform grid on the quarter domain. Based on our past experience, all computations of stability boundaries here are initiated at $\beta_o = 2$ (i.e., $W_o = 10$).

In Figure 4.1 we depict the computed wrinkling stability boundaries for thickness $h = 0.01$, for each of the four models NH, MR, S-VK and FvK presented in Section 2, cf. (9), (10), (13)$_{1,2}$, (14) and (15). Observe that the three finite-elasticity (membrane) models, NH, MR and S-VK, yield closed curves, outside of which there is no wrinkling; parameter values inside the closed boundary correspond to wrinkled states. On the other hand, the FvK stability boundary is open, to the left of which are non-wrinkling parameter values, with wrinkling to the right of the open curve. Already these results suggest that the FvK model predicts wrinkling for all aspect ratios above a certain threshold, the value of which is around $\beta = L/W = 1.25$.

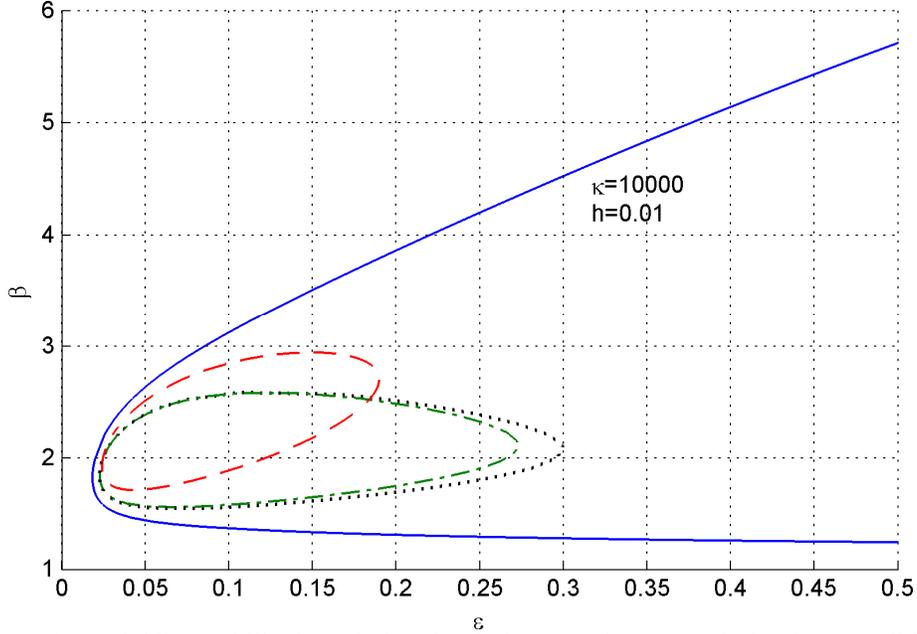

Figure 4.1 Wrinkling stability boundaries: dot-dash = NH, dot = MR, dash = S-VK, solid = FvK.

In conjunction with Figure 4.1, Figures (4.2a-d) depict the bifurcation diagrams (maximum wrinkling amplitude, max |w| vs. macroscopic strain, $\varepsilon$) for stable wrinkled states at various aspect ratios $\beta = L/W$, as indicated. Observe that the three finite-elasticity models predict the "birth and death" of isola-center bifurcations as the aspect ratio $\beta$ is varied. In particular, wrinkling occurs only for aspect ratios $\beta$ contained in a bounded interval, as was first identified in Healey, et. al. (2013) for the S-VK model. While the latter gives the correct qualitative behavior, it does not compare well with the predictions from the other two more accurate models NH and MR, the results of which compare well with each other. In particular, the bounded $\beta$-interval engendering wrinkling is more narrow for NH and NR, and it is nearly centered at $\beta = 2$: For the NH and NR models, we see from Figure 4.1 that an isola-center begins at about $\beta = 1.5$, after which wrinkling occurs, and terminates around $\beta = 2.6$, after which there is no wrinkling. Also, within most that same $\beta$ interval, the predicted maximum strain at which the wrinkles disappear is larger for NH and MR than that predicted by S-VK.

Observe that the FvK model always predicts a pitchfork bifurcation. As noted in our earlier work, this corresponds to an ever-increasing wrinkling amplitude as the macroscopic strain is increased, which is certainly unrealistic. Moreover, as $\beta$ approaches the apparent horizontal asymptote of the FvK stability boundary in Figure 4.1 or takes on very large values, the pitchforks predicted by the FvK model shown in Figures (4.2a,d) move out to extremely large values of the macroscopic strain. This is in contrast to the finite-elasticity models, which predict that (for appropriate aspect ratios in a bounded interval) wrinkling initiates, reaches a small maximum amplitude, diminishes and then disappears altogether, as the macroscopic strain is steadily increased.

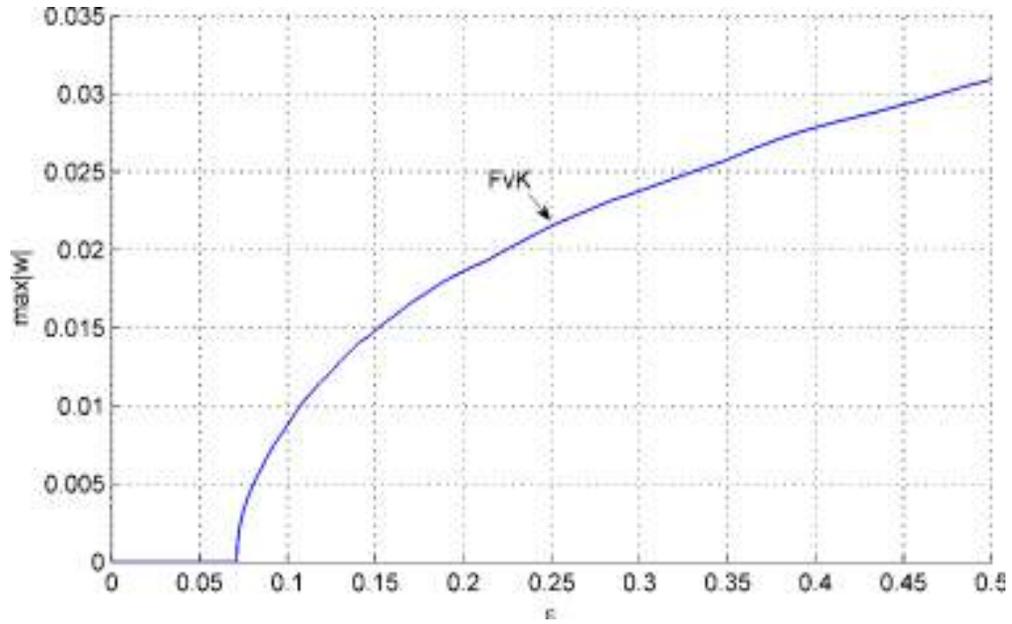

Figure 4.2(a) Bifurcation diagrams, $\max |w|$ vs. $\varepsilon$, $h = 0.01, \beta = 1.4$.

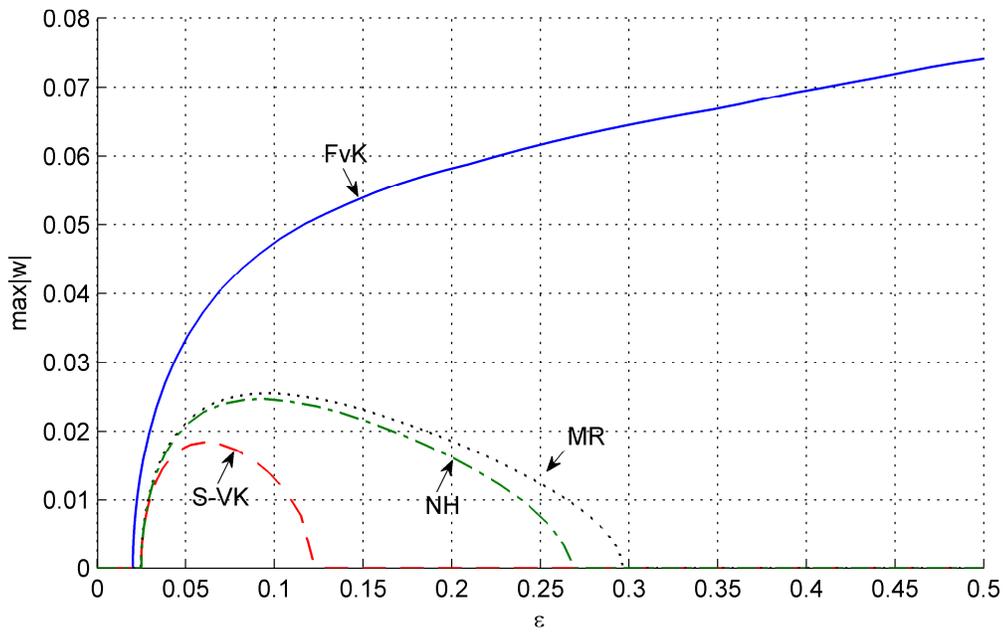

Figure 4.2(b) Bifurcation diagrams, $\max |w|$ vs. $\varepsilon$, $h = 0.01, \beta = 2$.

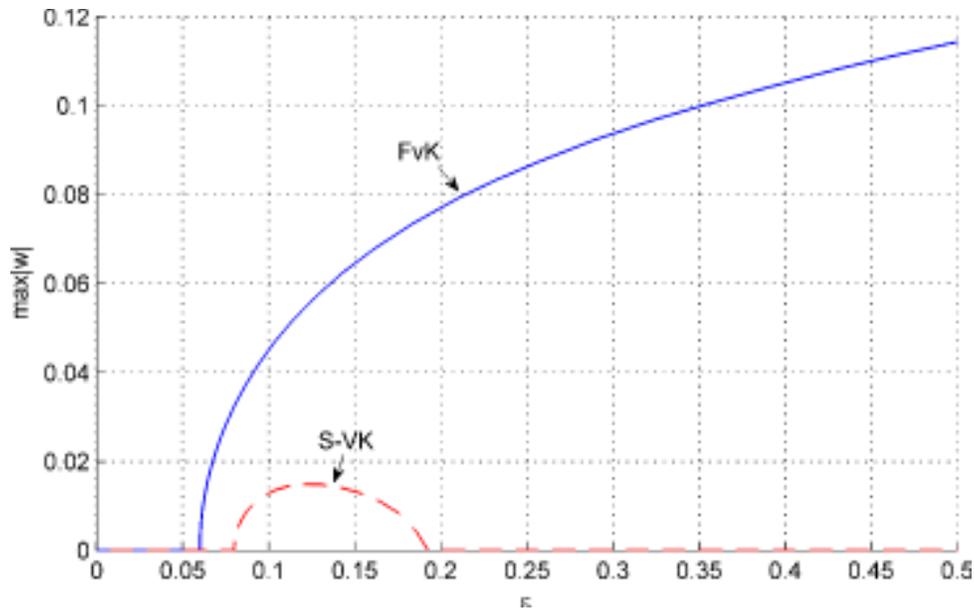

Figure 4.2(c) Bifurcation diagrams, $\max|w|$ vs. $\varepsilon$, $h = 0.01$, $\beta = 2.75$.

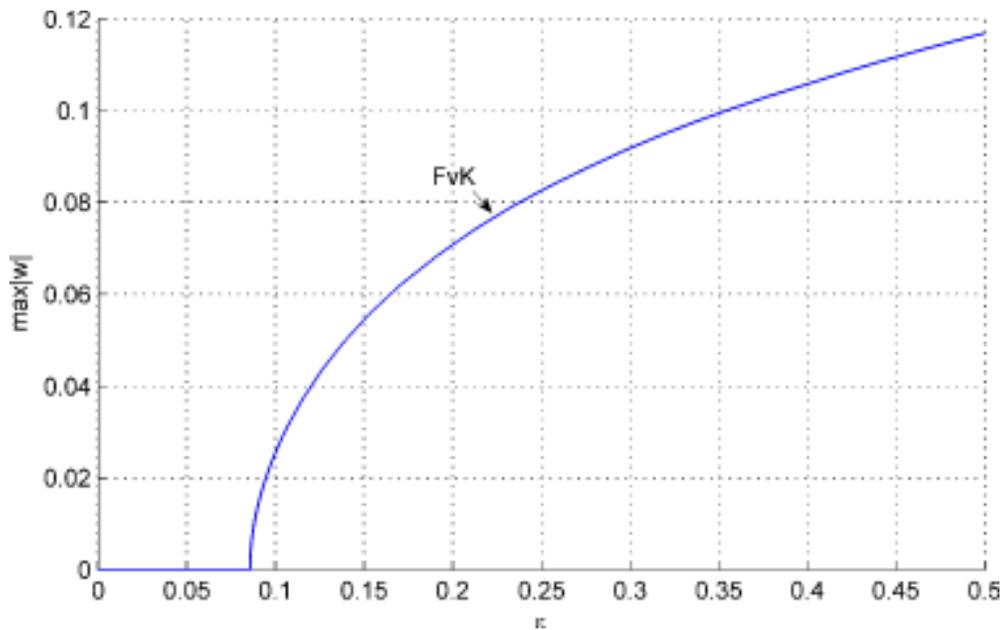

Figure 4.2(d) Bifurcation diagrams, $\max|w|$ vs. $\varepsilon$, $h = 0.01$, $\beta = 3$.

Figures 4.3(a-e) and 4.4(a-e) depict bifurcation diagrams for the two models NH and MR, each with aspect ratio $\beta = 2$ and thickness $h = 0.01$, accompanied by four corresponding wrinkled configurations, as marked on the respective bifurcation diagram.

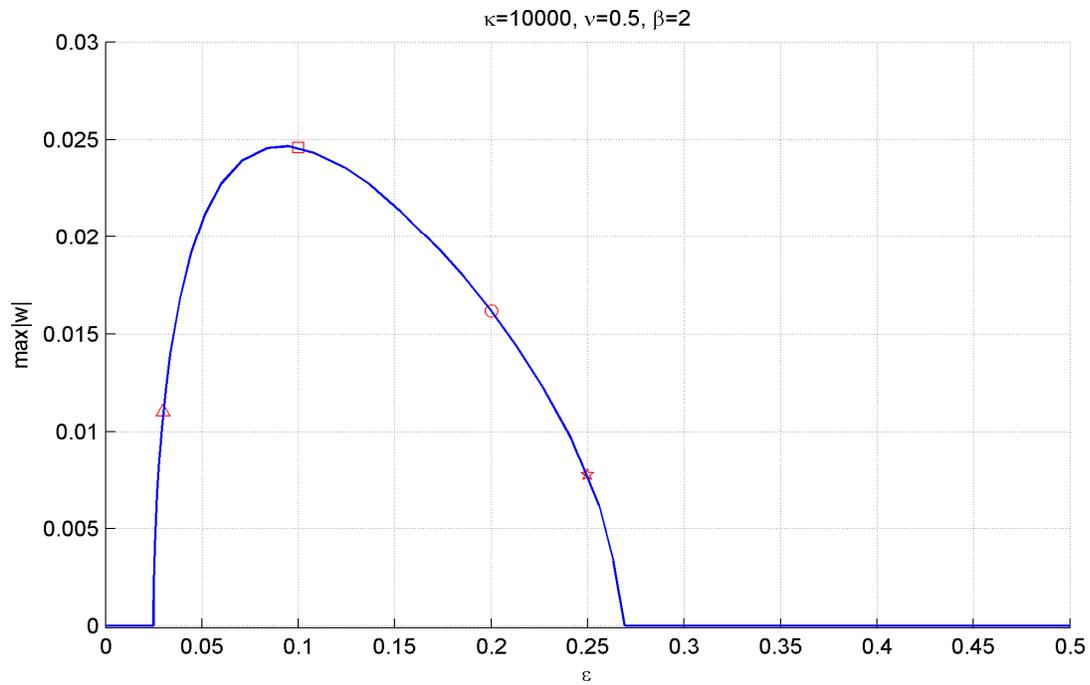

Figure 4.3(a) Bifurcation diagram for NH model: $\beta = 2, h = 0.01$.

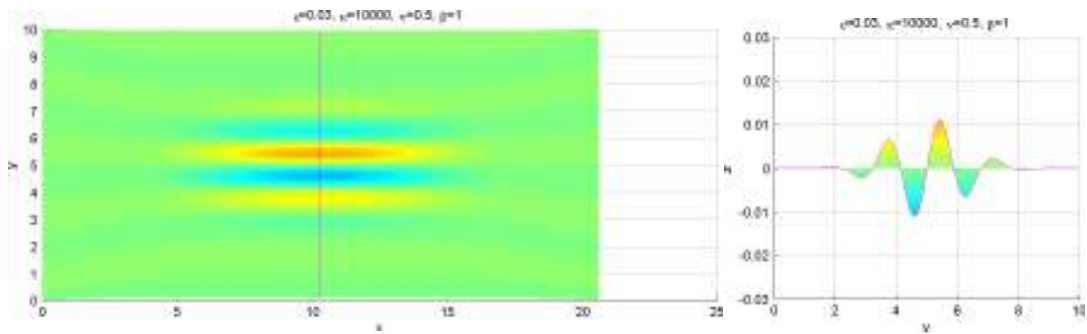

Figure 4.3(b) Wrinkled configuration: NH model, $\varepsilon = 0.03, \beta = 2, h = 0.01$.

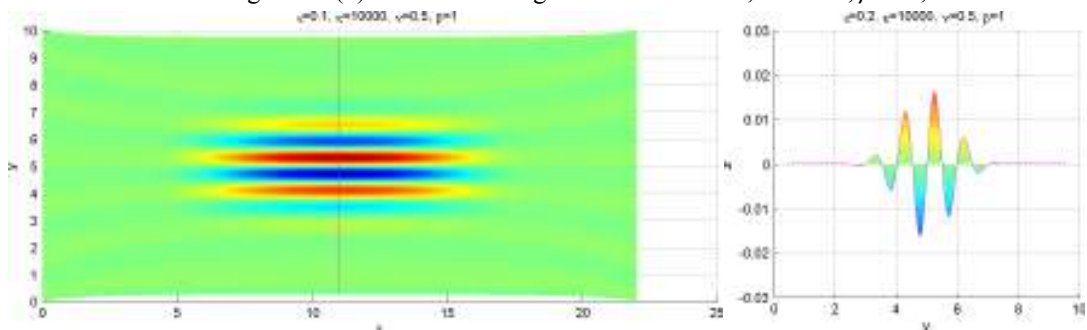

Figure 4.3(c) Wrinkled configuration: NH model, $\varepsilon = 0.1, \beta = 2, h = 0.01$.

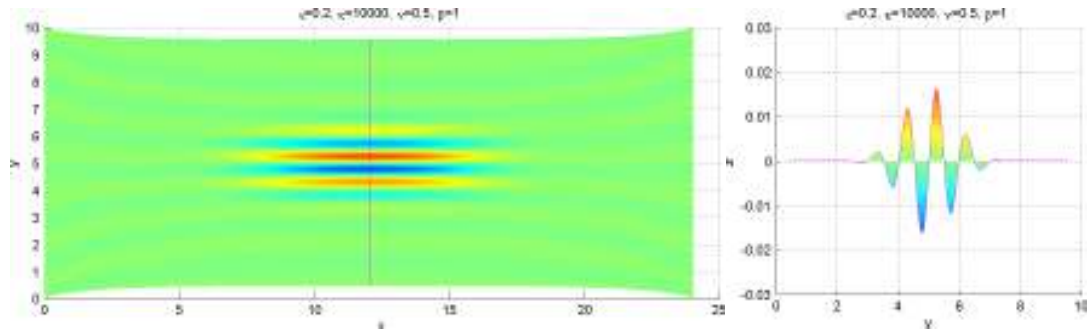

Figure 4.3(d) Wrinkled configuration: NH model, $\varepsilon = 0.2, \beta = 2, h = 0.01$.

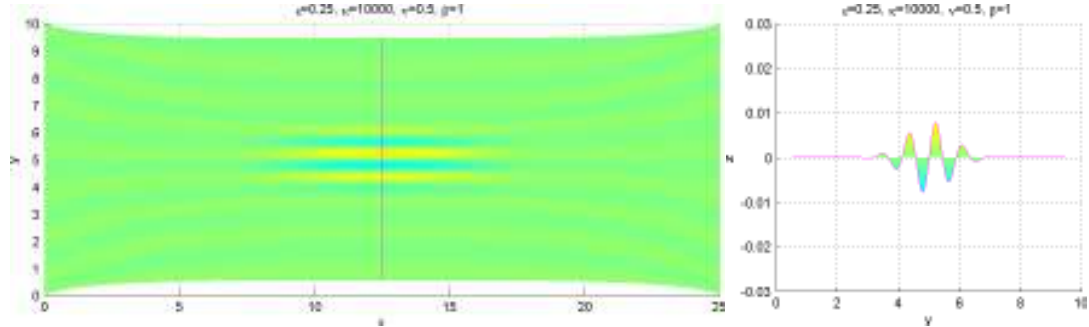

Figure 4.3(e) Wrinkled configuration: NH model, $\varepsilon = 0.25, \beta = 2, h = 0.01$.

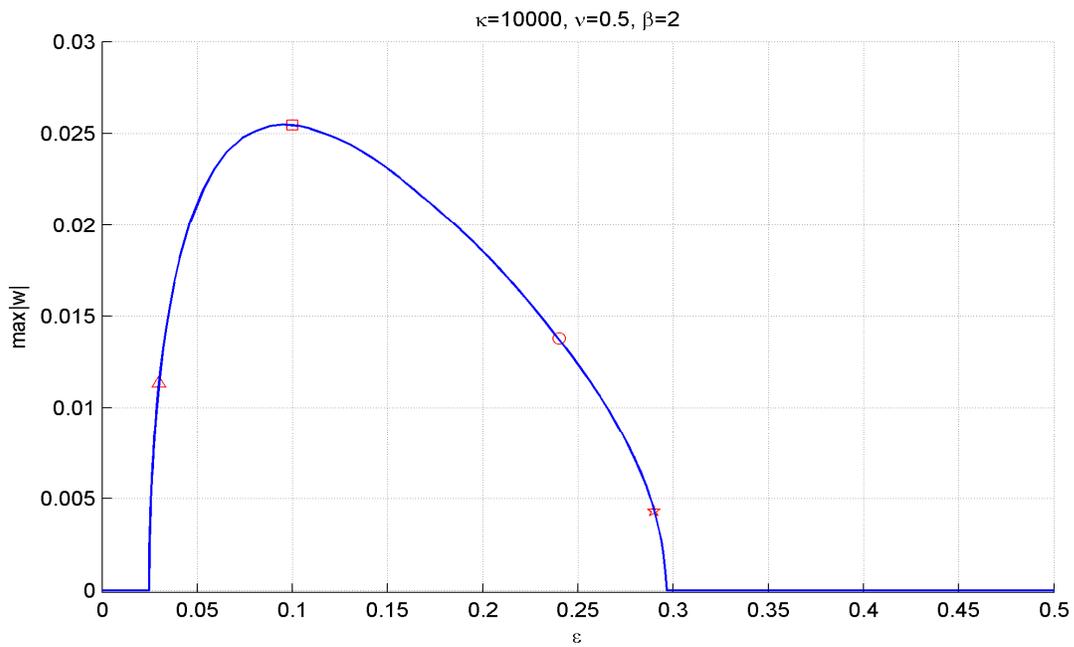

Figure 4.4(a) Bifurcation diagram for MR model: $\beta = 2, h = 0.01$.

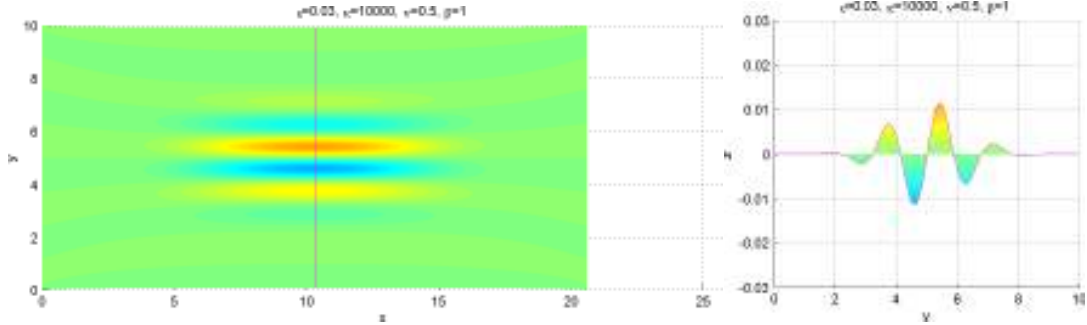
Figure 4.4(b) Wrinkled configuration: MR model, $\varepsilon = 0.03, \beta = 2, h = 0.01$.

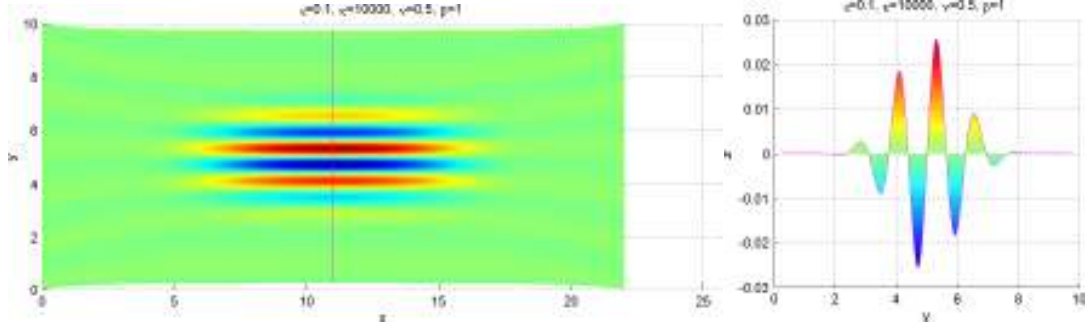
Figure 4.4(c) Wrinkled configuration: MR model, $\varepsilon = 0.1, \beta = 2, h = 0.01$.

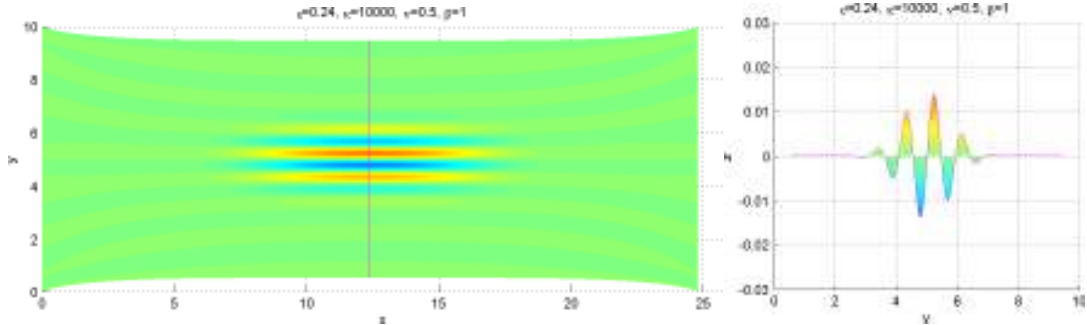
Figure 4.4(d) Wrinkled configuration: MR model, $\varepsilon = 0.24, \beta = 2, h = 0.01$.

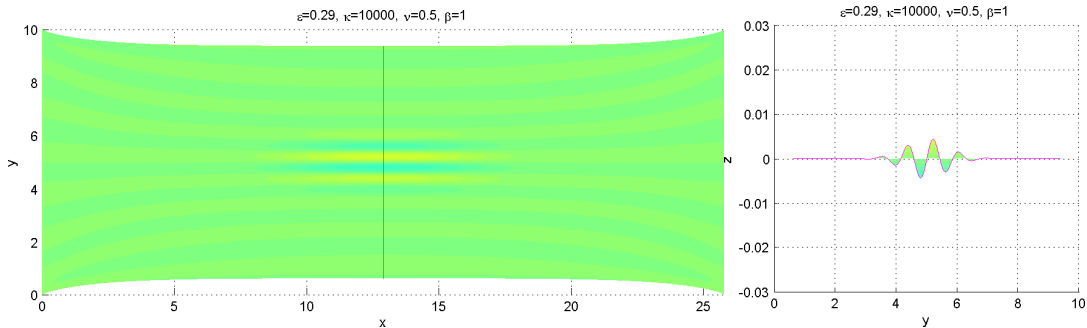
Figure 4.4(e) Wrinkled configuration: MR model, $\varepsilon = 0.29, \beta = 2, h = 0.01$.

In the following Figures 4.5-4.8, we present the same sequence of data as given above in Figures 4.1-4.4, but now for thickness $h = 0.005$. The overall trend is the same, although the wrinkling regions are now larger, as shown in Figure 4.5. Figures 4.6(a-d) complement Figure 4.5, and Figures 4.7 and 4.8 depict bifurcation diagrams along with associated configurations for the NH and MR models, respectively. All comments made above for thickness $h = 0.01$ are pertinent here as well, and are not repeated.

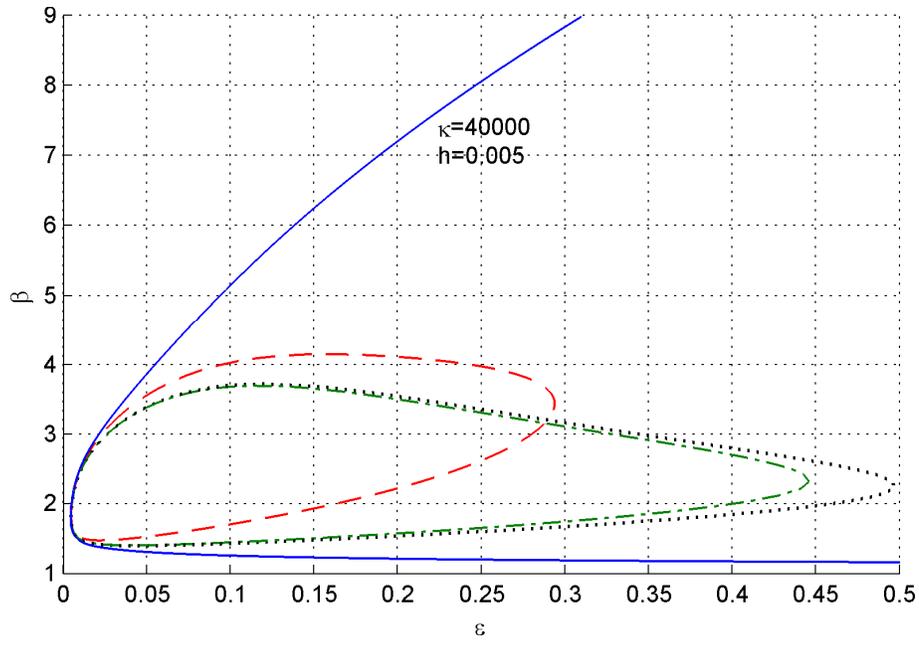

Figure 4.5 Wrinkling stability boundaries: dot-dash = NH, dot = MR, dash = S-VK, solid = FvK.

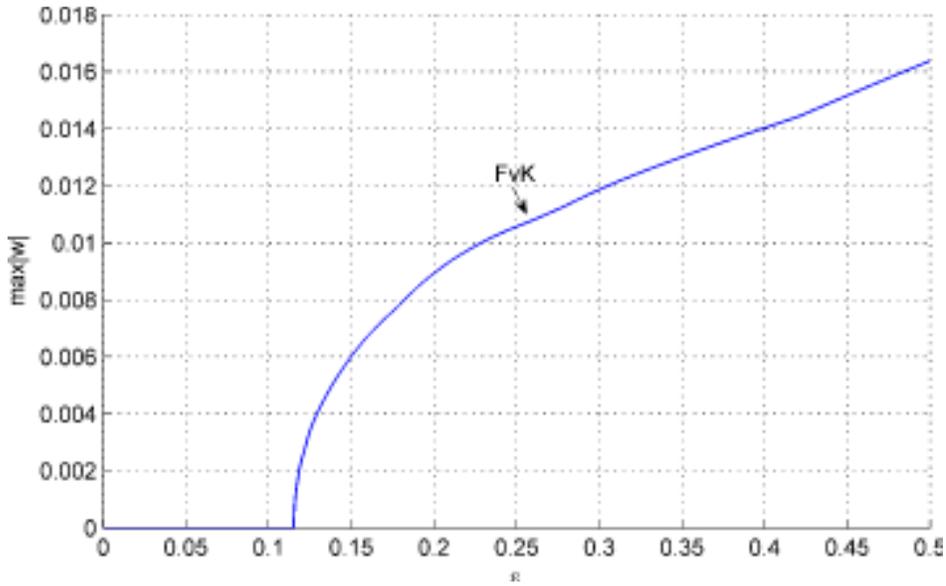

Figure 4.6(a) Bifurcation diagrams, $\max|w|$ vs. $\varepsilon$, $h = 0.005, \beta = 1.25$.

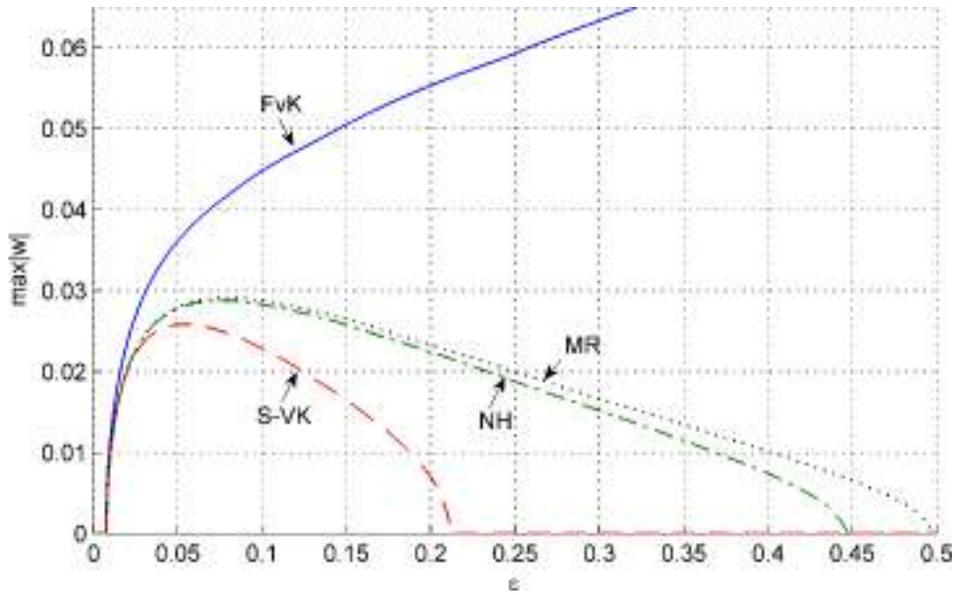

Figure 4.6(b) Bifurcation diagrams, max $|w|$ vs. $\varepsilon$, $h = 0.005$, $\beta = 2.3$.

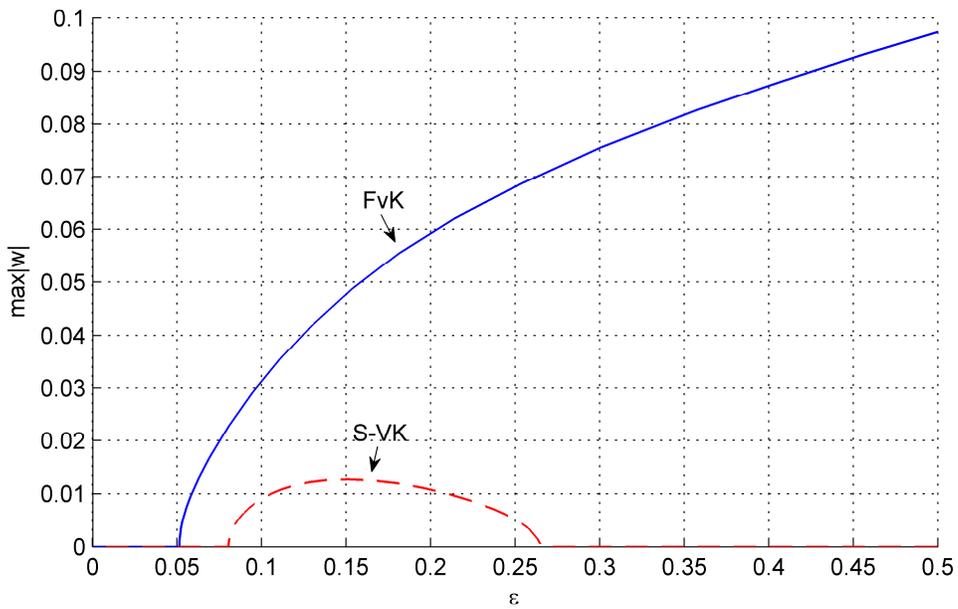

Figure 4.6(c) Bifurcation diagrams, max $|w|$ vs. $\varepsilon$, $h = 0.005$, $\beta = 3.9$.

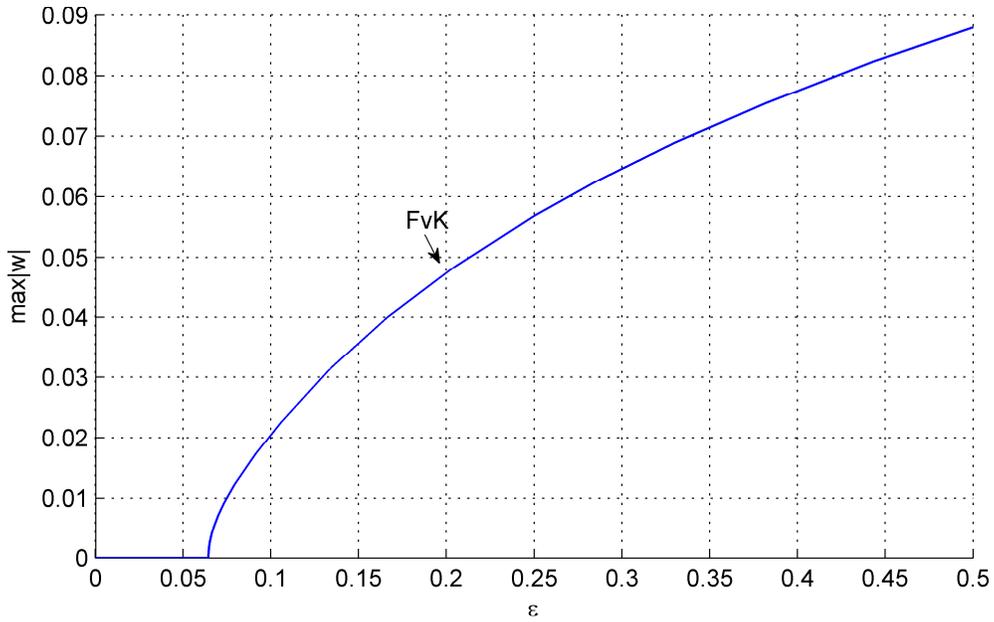

Figure 4.6(d) Bifurcation diagrams, max $|w|$ vs. $\varepsilon$, $h = 0.005, \beta = 4.25$.

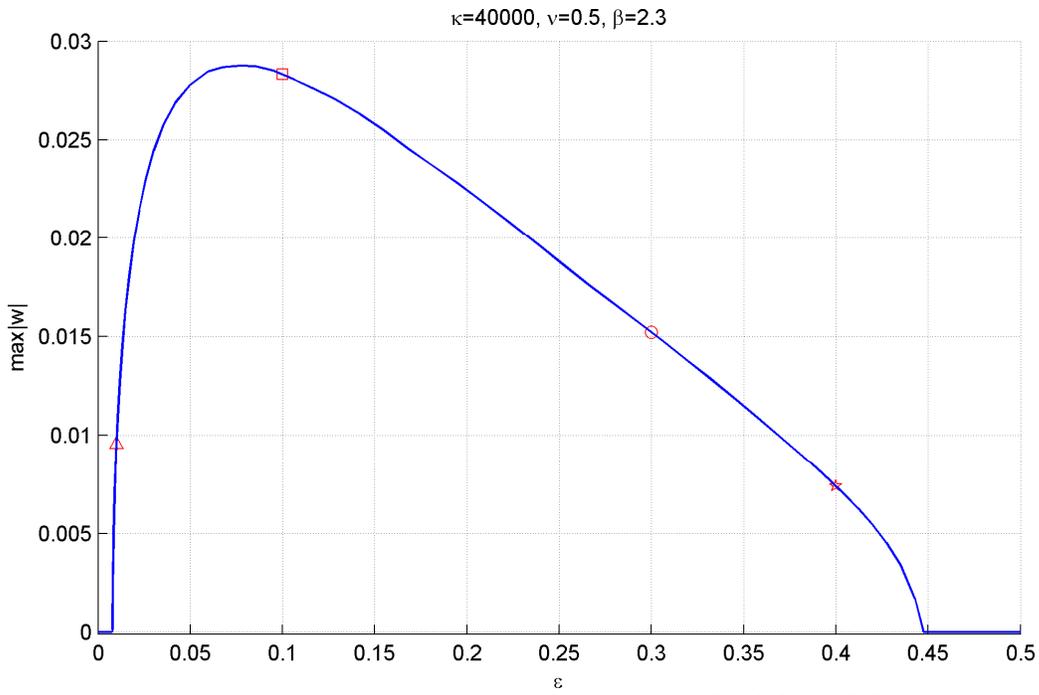

Figure 4.7(a) Bifurcation diagram for NH model: $\beta = 2.3, h = 0.005$.

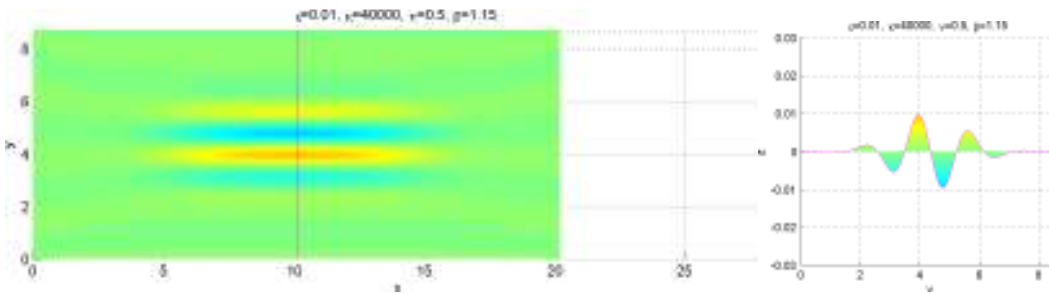

Figure 4.7(b) Wrinkled configuration: NH model, $\varepsilon = 0.01, \beta = 2.3, h = 0.005$.

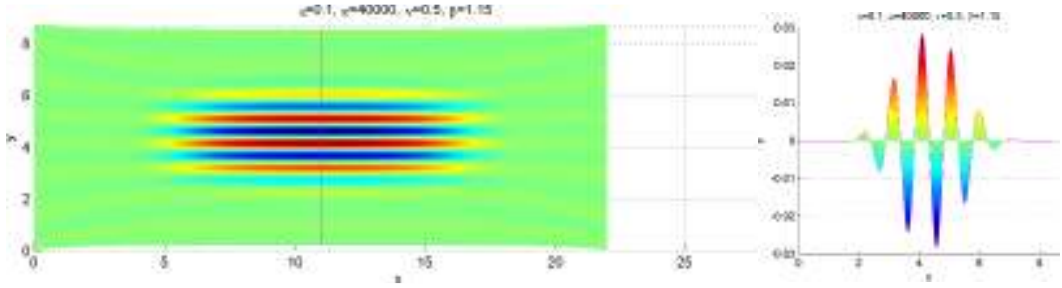

Figure 4.7(c) Wrinkled configuration: NH model, $\varepsilon = 0.1, \beta = 2.3, h = 0.005$.

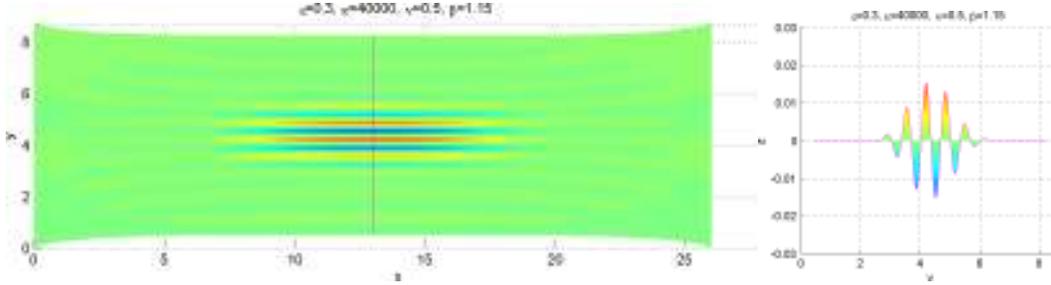

Figure 4.7(d) Wrinkled configuration: NH model, $\varepsilon = 0.3, \beta = 2.3, h = 0.005$.

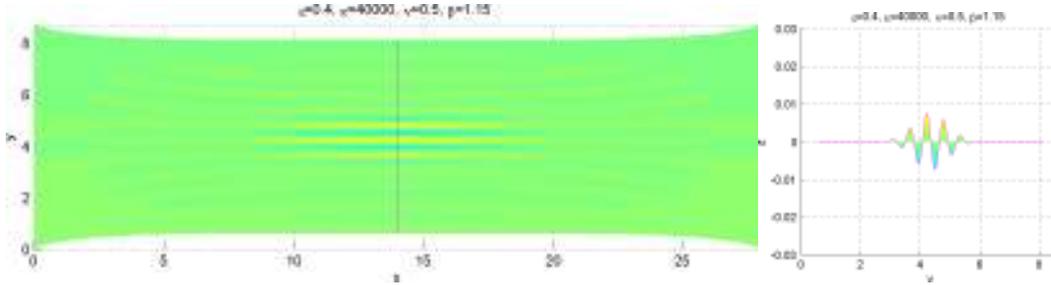

Figure 4.7(d) Wrinkled configuration: NH model, $\varepsilon = 0.4, \beta = 2.3, h = 0.005$.

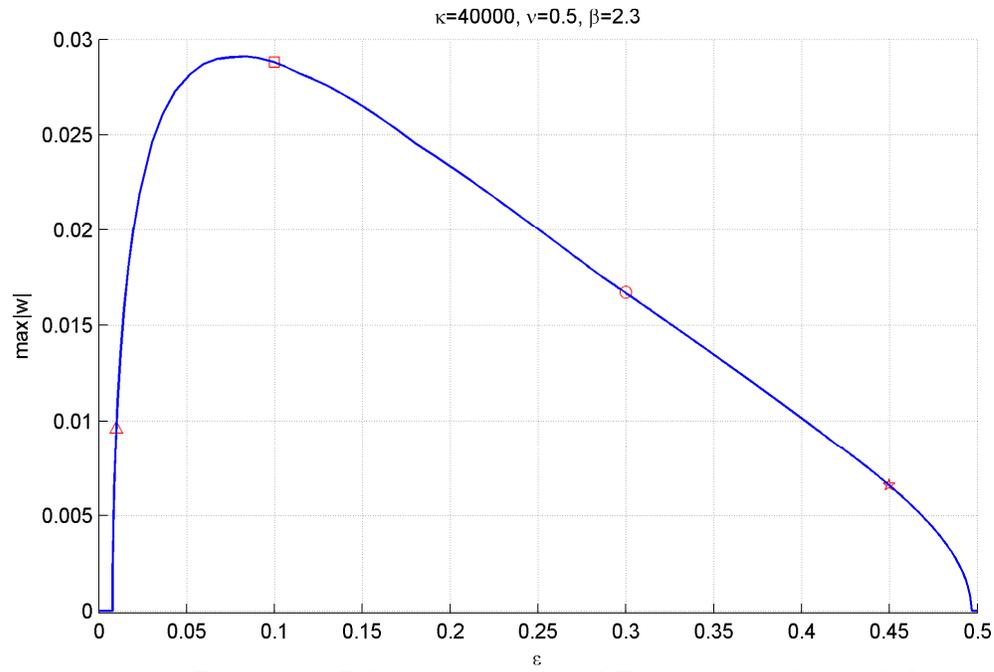

Figure 4.8(a) Bifurcation diagram for MR model: $\beta = 2.3, h = 0.005$.

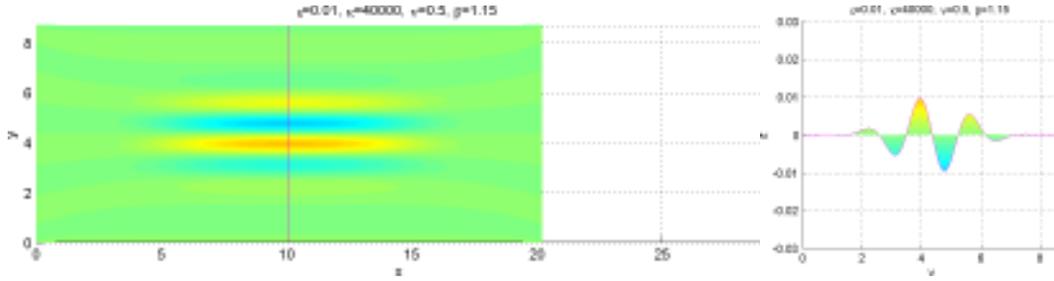
Figure 4.8(b) Wrinkled configuration: MR model, $\varepsilon = 0.01, \beta = 2.3, h = 0.005$.

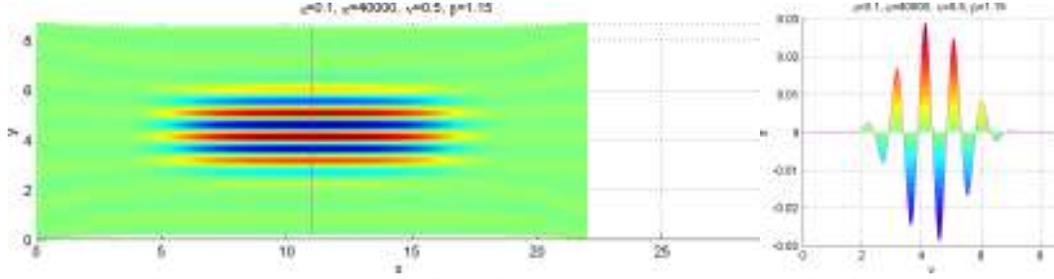
Figure 4.8(c) Wrinkled configuration: MR model, $\varepsilon = 0.1, \beta = 2.3, h = 0.005$.

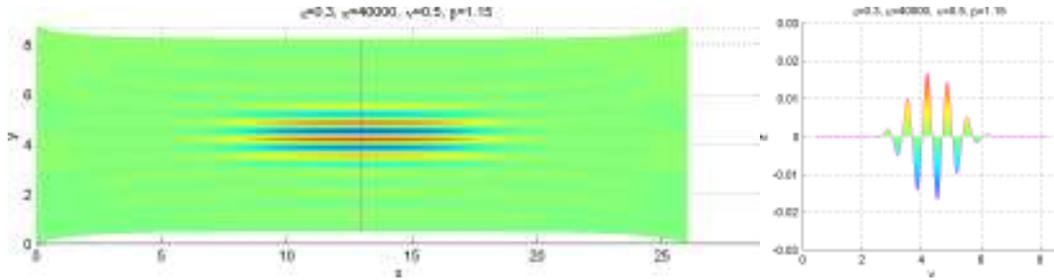
Figure 4.8(d) Wrinkled configuration: MR model, $\varepsilon = 0.3, \beta = 2.3, h = 0.005$.

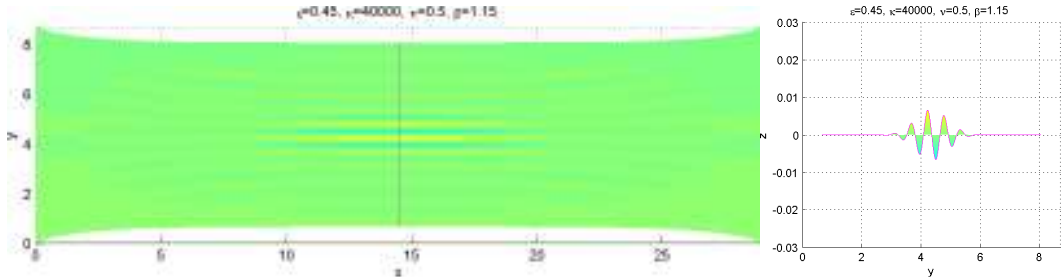
Figure 4.8(d) Wrinkled configuration: MR model, $\varepsilon = 0.45, \beta = 2.3, h = 0.005$.

## 5. Concluding Remarks

Our contribution to the wrinkling problem at hand is two-fold: (1) We bring our systematic global bifurcation/continuation approach to bear on this problem using the more accurate elastomer models NH and MR. (2) We provide an efficient strategy for the computation of stability boundaries (aspect ratio vs. macroscopic strain) and present them for four different models – NH, MR, S-VK, FvK. The computed boundaries reveal serious shortcomings of the FvK model and apparent inaccuracies of the S-VK model, the latter in comparison with the results for the more realistic models NH and MR.

The 2D membrane models S-VK, cf. (13)$_1$, and NH, cf. (9), are identical to those inherent in the models used by Taylor, et. al. (2013) and Nayyar, et. al. (2011), respectively. On the other hand, each of those works employs a more sophisticated bending model than the one used here: Taylor, et. al. (2013) use (14) without the simplification (15), while Nayyar, et. al. (2011) employ a geometrically exact, hyperelastic shell model from ABACUS for a neo-Hookean material, which is presumably more refined than (14). Nonetheless, we contend that (14), (15) – combined either with (9) or especially with (10) - is accurate for the problem at hand. First, recall that the 3D incompressible neo-Hookean and Mooney-Rivlin constitutive laws are based upon experiments for biaxially loaded rubber sheets; the latter is especially accurate in that setting, cf. Mooney (1940), Müller and Strehlow (2004). In that sense, our model MR here is superior. In particular, the inaccuracy of the S-VK model in predicting stability boundaries, the crucial membrane part of which is identical to that of the Koiter model, is clearly indicated in Section 4.

As for our linear bending model, we already pointed out in the Introduction that the wavelength of a typical wrinkle in this setting is two orders of magnitude greater than the maximum wrinkling amplitude, the latter of which is the same order as the fine thickness. Now the maximum magnitude of the gradient, viz., $|\nabla w|$, is approximately twice the wrinkling amplitude divided by a half of wavelength, which is the same order of magnitude as the wrinkling displacement. As in the FvK model, (15) is the (small-gradient) linearziation of the exact curvature tensor about the flat state. We further note that in the numerical implementation of the models employed in the two papers discussed above, the bending energy is multiplied – once and for all – by the square of the fine thickness (in comparison to the membrane energy). It seems clear that a geometrically exact bending energy is not important in this problem; the accuracy of the finite-elasticity membrane model is crucial. In any case, *we can say with certainty that a more sophisticated bending model will not alter the stability boundaries*; $(32)_1$ is the planar membrane problem, while $(32)_2$ involves the *linearized* bending operator.

Existence theory for this problem is difficult – at least in terms of providing justification for bifurcation analysis. The situation is similar to that of classical nonlinear elasticity (and the analysis of elliptic partial differential equations, in general): Mixed boundary conditions and/or corners preclude the smoothness of solutions required to justify rigorous techniques of bifurcation theory. However, direct energy minimization is a more natural and promising avenue, which we intend to pursue elsewhere.

**Acknowledgements:** The work of TJH was supported in part by the National Science Foundation through grant DMS-1312377. The work of QL was supported in part by the Science Fund for Distinguished Young Scholars of Chongqing (cstc2013jcyjjq40001). Each of these is gratefully acknowledged. We also acknowledge fruitful discussions with Andras Sipos and Kyung-Suk Kim concerning this work.